\documentclass[aps, prx, reprint, longbibliography]{revtex4-1}
\usepackage{amsmath}
\usepackage[pdftex]{graphicx}
\usepackage{graphicx}
\usepackage{amsmath}
\usepackage{amssymb}
\usepackage{url}
\usepackage{bm}		
\usepackage{color}	
\usepackage{textcomp}				

\begin{document}


\title{Exact solution to the steady-state dynamics of a periodically-modulated resonator} 



\author{Momchil Minkov}
\email[]{mminkov@stanford.edu}

\author{Yu Shi}

\author{Shanhui Fan}
\email[]{shanhui@stanford.edu}
\affiliation{Department of Electrical Engineering, and Ginzton Laboratory, Stanford University, Stanford, CA 94305, USA}

\date{\today}

\begin{abstract}
We provide an analytic solution to the coupled-mode equations describing the steady-state of a single periodically-modulated optical resonator driven by a monochromatic input. The phenomenology of this system was qualitatively understood only in the adiabatic limit, i.e. for low modulation speed. However, both in and out of this regime, we find highly non-trivial effects for specific parameters of the modulation. For example, we show complete suppression of the transmission even with zero detuning between the input and the static resonator frequency. We also demonstrate the possibility for complete, lossless frequency conversion of the input into the side-band frequencies, as well as for optimizing the transmitted signal towards a given target temporal waveform. The analytic results are validated by first-principle simulations.
\end{abstract}


\maketitle

\section{Introduction} 

Silicon photonics has taken a central role in communication technologies, and is becoming competitive to the conventional electronic signal transport on shorter and shorter length-scales \cite{Priolo2014, Miller2016}. We have in fact reached the point at which \textit{chip-scale} photonic technologies are viable candidates for on-chip interconnect applications \cite{Miller2016}. The electro-optic modulator \cite{Reed2010, Liu2004} is one of the most important components in silicon photonics, and CMOS-compatible, micrometer-scale devices based on a modulated cavity resonance have been a central focus of research \cite{Xu2005, Xu2007, Tanabe2009, Gardes2009, Koos2009, Wuelbern2009, Sacher2008, Sacher2009, Sandhu2012, Sacher2013, Manipatruni2010, Alloatti2012, Yu2014, Timurdogan2014}. Typically in these systems, a local refractive index change of the silicon results in a change of the transmission through the cavity. This is intuitively understood in the adiabatic limit, in which the modulation happens on a time-scale that is much slower than the one given by the photon life-time, but the phenomenology is in general much richer \cite{Sacher2008, Sacher2009, Sandhu2012, Sacher2013}. Here, we solve exactly the steady-state dynamics of a cavity with a peridocially-modulated resonance frequency. We provide a quantitative definition of the adiabatic regime, and find highly non-intuitive effects outside of it. The applications include transmission switching, lossless frequency conversion, and  signal manipulation, and are thus relevant to the broad field of photonic communication. Furthermore, systems of periodically-modulated, \textit{coupled} resonator modes have recently been shown to break reciprocity, which can be used for non-magnetic photonic isolation \cite{Hafezi2012, Sounas2014}, and even for photonic topological insulators \cite{Fang2012, Yuan2016, Minkov2016}. The approach we take in this work opens up a perspective to study these systems analytically, beyond the commonly employed approximations.

\section{Theory}

\begin{figure}
\centering
\includegraphics[width = 0.4\textwidth, trim = 0in 0in 0in 0in, clip = true]{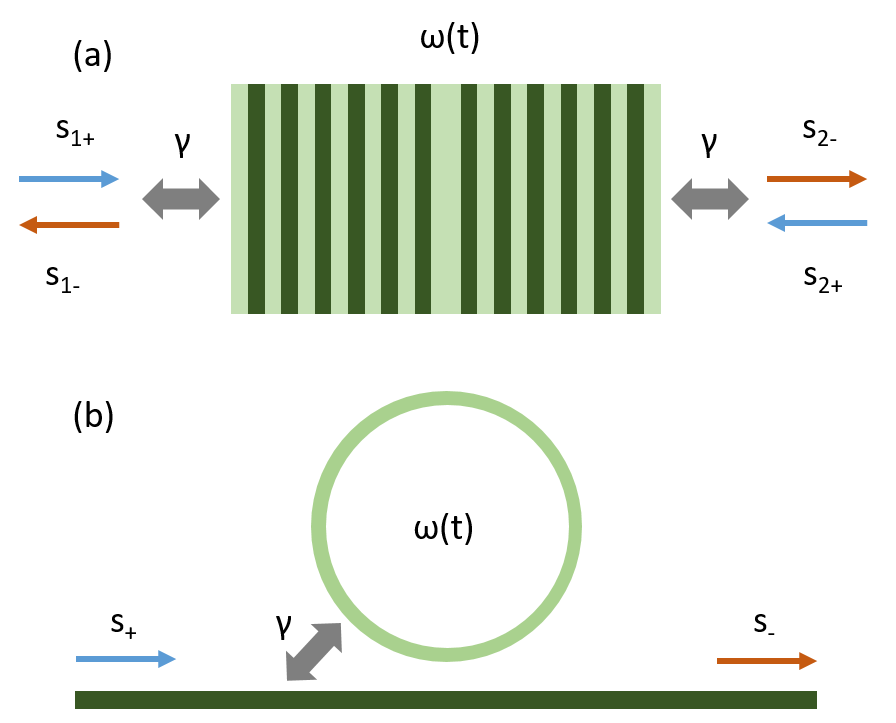}%
 \caption{Schematic of (a): a DBR cavity with an input/output port on each side, and (b): a microring cavity coupled to a single input/output port.}
\label{fig1}
\end{figure}

We study two optical cavity configurations relevant to chip-scale technologies (Fig. \ref{fig1}). The first one is a cavity coupled to two input/output ports, schematically represented as a distributed Bragg reflector (DBR) cavity in Fig. \ref{fig1}(a). Following Refs. \cite{Haus1984, Fan2003}, the coupled-mode (CM) equations for this system read
\begin{align}
&\frac{\mathrm{d} \alpha}{\mathrm{d} t} = (i(\omega_0 + \omega(t)) - \gamma)\alpha + \sqrt{\gamma} s_{1+}, \label{eqn:starting1} \\  \label{eqn:starting2}
&s_{1-} = -s_{1+} + \sqrt{\gamma} \alpha(t); \quad s_{2-} = \sqrt{\gamma} \alpha(t),
\end{align}
with $|\alpha|^2$ representing the electromagnetic energy inside the resonator, while $s_{j+}$ and $s_{j-}$ are the input and output amplitudes in the $j$-th port, respectively. $|s_{j+}|^2$ and $|s_{j-}|^2$ correspond to input and output power. The resonance has a decay rate of $\gamma$. Here we assume that the resonance decays entirely through the coupling to the ports, and moreover the decay rates to the two ports are equal. The resonance frequency in the absence of modulation is $\omega_0$, and $\omega(t)$ is the time-dependent modulation. The second configuration involves a single in/out port and is relevant for example to the case of a micro-ring resonator side-coupled to a waveguide (Fig. \ref{fig1}(b)). For such a configuration, we have
\begin{align}
\label{eqn:mr1}
&\frac{\mathrm{d} \alpha}{\mathrm{d} t} = (i(\omega_0 + \omega(t)) - \gamma/2)\alpha_i + i\sqrt{\gamma} s_{+}, \\ 
&s_{-} = s_{+} + i\sqrt{\gamma} \alpha(t). \label{eqn:mr2}
\end{align}
We note that the eqs. (\ref{eqn:starting1}-\ref{eqn:starting2}) describing panel (a) are also relevant to a micro-ring resonator coupled to \textit{two} waveguides, one on each side. The crucial difference between the two systems in Fig. \ref{fig1} is the fact that in the one of panel (b), the time-integrated transmitted power is always equal to the input power, since there is only one output port and energy conservation holds. In contrast, in panel (a), the power can be arbitrarily split between the transmission channel ($s_{2-}$) and the reflection channel ($s_{1-}$).

Next, we solve analytically eqs. (\ref{eqn:starting1}-\ref{eqn:starting2}) for the case when $s_{1+} = p_0 \exp(i\omega_p t)$ and the frequency modulation is given by $\omega(t) = A_0 \cos(\Omega t)$. Under the gauge transformation 
\begin{equation}
\beta(t) = \alpha(t) \times \exp\left(-i\omega_p t -i \int_0^t \omega(t') \mathrm{d}t'\right),
\end{equation} 
the equation of motion for the amplitude $\beta$ reads
\begin{equation}
\frac{\mathrm{d} \beta}{\mathrm{d} t} = (- i\Delta \omega - \gamma)\beta + \sqrt{\gamma} p_0 e^{-i \frac{A_0}{\Omega}\sin(\Omega t)},
\end{equation}
where we labeled by $\Delta \omega = \omega_p - \omega_0$ the detuning between the source frequency and the un-modulated resonance frequency. The complex phase dependence of the source term can be simplified using the Jacobi-Anger expansion:
\begin{equation}
\frac{\mathrm{d} \beta}{\mathrm{d} t} = (-i\Delta \omega - \gamma)\beta + \sqrt{\gamma} p_0 \sum_n \mathcal{J}_n\left(\frac{A_0}{\Omega}\right) e^{-in\Omega t}.
\end{equation}
This equation describes a cavity at a \textit{fixed} resonance frequency driven by infinitely many sources, one at each frequency $n \Omega$ with amplitude $p_0 \mathcal{J}_n(A_0/\Omega)$ for every integer $n$.

Next, we make an Ansatz for the steady-state solution of the system, namely we look for a solution $2\pi/\Omega$-periodic in time, such that 
\begin{equation}
\beta(t) = \sum_k \beta_k e^{-ik\Omega t} 
\end{equation}
with $\beta_k$ constants that do not depend on time. We have checked that this is justified by the full dynamical solution to eqs. (\ref{eqn:starting1}-\ref{eqn:starting2}), which converges to such a stationary state at times larger than $1/\gamma$. With this form of $\beta$, we have
\begin{equation}
\sum_k (ik \Omega - i\Delta \omega - \gamma)\beta_k e^{-ik \Omega t} = - \sqrt{\gamma} p_0 \sum_n \mathcal{J}_n\left(\frac{A_0}{\Omega}\right) e^{-in\Omega t}.
\end{equation}
This can only be satisfied at all times if the sums are equalized term by term, yielding
\begin{equation}
\beta_k = - \sqrt{\gamma} p_0 \mathcal{J}_k\left(\frac{A_0}{\Omega}\right) \frac{1}{ik \Omega - i\Delta \omega - \gamma}.
\end{equation}

The result for $\beta_k$ is the same as the steady-state amplitude of a cavity at frequency $\Delta \omega$ pumped by an external field of amplitude $p_0 \mathcal{J}_k\left(\frac{A_0}{\Omega}\right)$ and frequency $k\Omega$. To compute the power transmitted into the second port, we first return to the starting gauge, using once again the Jacobi-Anger expansion:
\begin{align}
\nonumber \alpha(t) &= \sum_k \beta_k e^{i(\omega_p - k\Omega)t} \times \sum_n \mathcal{J}_{n} \left(\frac{A_0}{\Omega}\right) e^{in\Omega t}  \\ &= \sum_n \left(\sum_k \mathcal{J}_{n+k} \left(\frac{A_0}{\Omega}\right) \beta_k\right) e^{i(\omega_p + n\Omega)t}.
\end{align}
We can thus write the transmitted amplitude as
\begin{align}
& s_{2-}(t) = p_0 e^{i\omega_p t} \sum_n s_n e^{i n\Omega t} , \label{eqn:s2-}
\\ \label{eqn:s_n}
& s_n = -\sum_k \mathcal{J}_{n+k} \left(\frac{A_0}{\Omega}\right) \mathcal{J}_{k} \left(\frac{A_0}{\Omega}\right) \frac{\gamma}{ik \Omega - i\Delta \omega - \gamma}.
\end{align}
This result is \textit{exact} for the steady-state. The transmittivity spectrum has a component at every side-band, and the amplitude at the side-band with frequency $n\Omega + \omega_p$ consists of the sum of resonant contributions at $k\Omega - \Delta \omega$, with appropriate weights. The normalized transmitted power is also $2\pi/\Omega$-periodic, and can be computed as
\begin{align}
T(t) = \frac{|s_{2-}|^2}{p_0^2} = \sum_n T_n e^{i n\Omega t}; \quad \quad T_n = \sum_m s_m^* s_{n+m}.
\label{eqn:trans}
\end{align}
For the micro-ring case of eqs. (\ref{eqn:mr1}-\ref{eqn:mr2}), we find the same expression for $s_-$ as eq. (\ref{eqn:s2-}), but with the Fourier amplitudes given by
\begin{align}
s_n = \delta_{n0} + \sum_k \mathcal{J}_{n+k} \left(\frac{A_0}{\Omega}\right) \mathcal{J}_{k} \left(\frac{A_0}{\Omega}\right) \frac{\gamma}{ik \Omega - i\Delta \omega - \gamma/2},
\label{eqn:trans_mr}
\end{align}
and the transmitted power can also be computed as in eq. (\ref{eqn:trans}).

The derivation above can be generalized for an arbitrary time-periodic modulation of the cavity frequency. As an illustration, we consider
\begin{equation}
\omega(t) = A_1 \cos(\Omega t) + A_2 \cos(2\Omega t + \phi_m).
\end{equation}
Going through the same procedure as above, the coefficients $\beta_k$ in the DBR case are found to be
\begin{equation}
\beta_k = - \frac{\sqrt{\gamma} p_0}{ik \Omega - i\Delta \omega - \gamma} \sum_q \mathcal{J}_{k-2q}\left(\frac{A_1}{\Omega}\right) \mathcal{J}_{q}\left(\frac{A_2}{\Omega}\right) e^{iq\phi_m},
\label{eqn:beta_2m}
\end{equation}
while the expression for the $n$-th spectral component of the transmitted amplitude
\begin{equation}
s_n = \sum_{p, k} \beta_k \mathcal{J}_{n + k-2p}\left(\frac{A_1}{\Omega}\right) \mathcal{J}_{p}\left(\frac{A_2}{\Omega}\right) e^{ip\phi_m}.
\label{eqn:sn_2m}
\end{equation}
In general, for any arbitrary, time-periodic modulation, one can perform a Fourier-series expansion $\omega(t) = \sum_n A_n \cos(n\Omega t + \phi_n)$ and then compute all the relevant spectral components. Every extra higher harmonic term in the Fourier expansion of $\omega(t)$ yields an additional summation in both eqs. (\ref{eqn:beta_2m}) and (\ref{eqn:sn_2m}).

\section{Limiting cases}

The theory presented so far is exact for all possible parameters of the modulation. To illustrate the results better, in this Section we focus on two limiting cases, in which simplifying approximations can be made. 

\subsection{Adiabatic limit}
We start our discussion with the most intuitive, adiabatic regime. We focus on the DBR cavity of Fig. \ref{fig1}(a), but the same line of argument can be followed for the micro-ring case, and in fact for any modulated cavity with an arbitrary time-dependence of the resonance frequency.

\begin{figure}[t]
\centering
\includegraphics[width = 0.46\textwidth, trim = 0in 0in 0in 0in, clip = true]{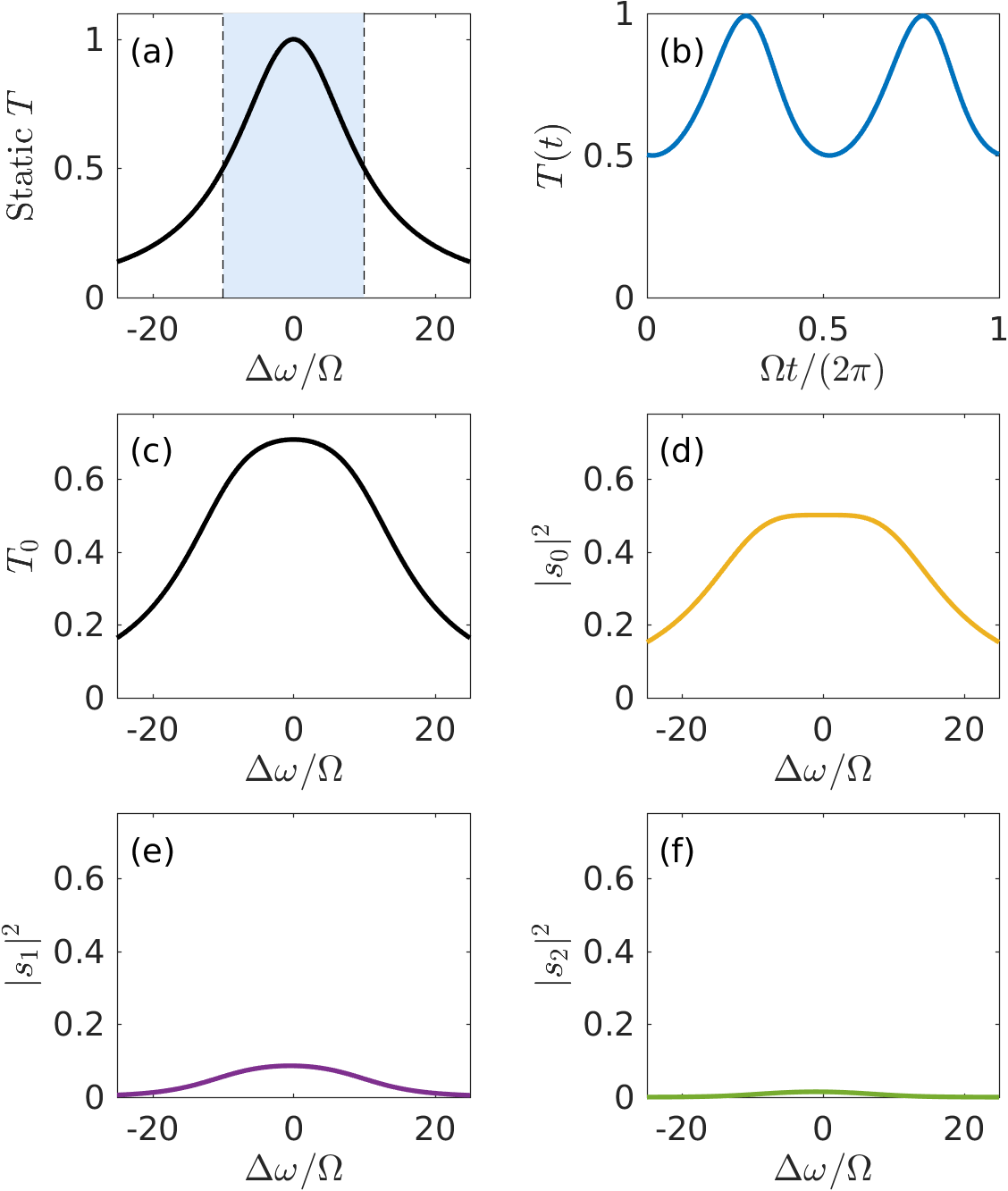}%
\caption{(a): Transmission vs. detuning for a static DBR cavity with $\gamma/\Omega = 10$. (b): Time-dependence of the transmission through the cavity of panel (a) under modulation with $A_0/\Omega = 10$ and $\Delta \omega = 0$. In this case the instantaneous resonant frequency of the cavity oscillates inside the blue shaded region of panel (a). (c): Time-averaged transmission through the modulated cavity as a function of $\Delta \omega$. (d)-(f): The power amplitudes in the zero-th, first, and second side-bands, respectively.}
\label{fig:adiabatic}
\end{figure}

When the dynamic modulation is sufficiently slow, we expect that the time-dependence of all physical quantities can be computed using an effective static frequency defined by $\omega(t)$ at every given $t$. More precisely, for example for the steady-state transmission we expect to write $T(t) = T(\omega(t))$, where 
\begin{equation}
T(\omega(t)) =  \left|\frac{\gamma}{i(\omega_0 + \omega(t) - \omega_p) - \gamma} \right|^2
\label{eqn:trans_ad}
\end{equation}
is the transmission of a static cavity at a resonance frequency $\omega_0 + \omega(t)$. Most numerical designs of resonant modulators implicitly assume this approximation \cite{Reed2010}. Under what condition should the system be in this regime, however, has not been previously discussed rigorously. Here, we derive self-consistently a condition that quantifies the range of parameters for which the system can be in this adiabatic regime. The following Ansatz,
\begin{equation}
\alpha(t) = \frac{\sqrt{\gamma} p_0}{i(\omega_0 +  \omega(t) - \omega_p) - \gamma},
\label{eqn:ad_alpha}
\end{equation}
can only be a steady-state solution to eq. (\ref{eqn:starting1}) if the left hand-side ($\mathrm{d} \alpha(t)/\mathrm{d} t$) is negligible with respect to the right hand-side. This imposes
\begin{equation}
\left|\frac{\sqrt{\gamma} p_0}{(i(\omega_0 + \omega(t) - \omega_p) - \gamma)^2} \frac{\mathrm{d} \omega(t)}{\mathrm{d} t} \right| \ll \sqrt{\gamma}p_0,
\end{equation}
or simply
\begin{equation}
\left| \frac{\mathrm{d} \omega(t)}{\mathrm{d} t} \right| \ll |(i(\omega_0 + \omega(t) - \omega_p) - \gamma)^2 |.
\end{equation}
For the particular case of cosine modulation, this reduces to
\begin{equation}
A_0 \Omega |\sin(\Omega t)| \ll |(i(-\Delta \omega + A_0 \cos(\Omega t)) - \gamma)^2|.
\label{eqn:ad_cos}
\end{equation}
The LHS of eq. (\ref{eqn:ad_cos}) is largest for $\Omega t = (2n+1)\pi/2$, $n \in \mathcal{Z}$, while the RHS is smallest at those same times, and for $\Delta \omega = 0$. Thus, for the evolution to be adiabatic at all times throughout the cycle and for all input frequencies, we finally obtain the condition $A_0 \Omega \ll \gamma^2$, or, in the units we use in this paper,
\begin{equation}
\frac{A_0}{\Omega} \ll \left(\frac{\gamma}{\Omega}\right)^2.
\label{eqn:ad_cond}
\end{equation}
It is important to note that the adiabatic limit is not simply defined by a small $\Omega$, but by an interplay of all three parameters $\Omega$, $A_0$, and $\gamma$. 

The adiabatic regime is illustrated in Fig. \ref{fig:adiabatic}, for a DBR cavity with $\gamma/\Omega = 10$. In panel (a), we plot the transmission vs. $\Delta \omega$ for the un-modulated cavity. In panel (b), we plot the time-dependent transmission for the same cavity, including a modulation with $A_0/\Omega = 10$ and detuning $\Delta \omega = 0$, as obtained using the exact result. The resonant frequency $\omega(t)$ oscillates within the blue shaded region of panel (a). Note that, for these parameters, the condition of eq. (\ref{eqn:ad_cond}) is satisfied, and the time-dependence of the transmission is very close to the prediction of the approximate result of eq. (\ref{eqn:trans_ad}). In panel (c), we plot the 0-th Fourier component $T_0$ of the transmission of the modulated cavity (see eq. (\ref{eqn:trans})). This is simply the normalized transmission integrated over a modulation cycle, and is also equal to $\sum |s_n|^2$. In panels (d)-(f), we show the transmitted power components $|s_0|^2$, $|s_1|^2$, and $|s_2|^2$, respectively. In general, following eq. (\ref{eqn:s_n}), one can show that
\begin{equation}
s_{-n}(\Delta \omega) = (-1)^n s_n^*(-\Delta \omega).
\label{eqn:s_-n}
\end{equation}
Thus, for all $n$ we have $|s_{-n}(\Delta \omega)|^2 = |s_{n}(-\Delta \omega)|^2$. In this adiabatic limit and under the cosine modulation, we can further show through the Fourier transform of eq. (\ref{eqn:ad_alpha}) that $|s_{n}(\Delta \omega)|^2 = |s_{-n}(\Delta \omega)|^2$. In Fig. \ref{fig:adiabatic}(c)-(f), this is manifested in the fact that the plots are symmetric with respect to $\Delta \omega = 0$, but this is not the case outside the adiabatic limit, as we will show below. An interesting aspect of Fig. 2(c)-(f) is that the transmission shows a flat-top rather than a Lorentzian-like lineshape. This flat-top transmission is characteristic of the `intermediate' adiabatic regime, when $A_0 \sim \gamma$. This is in contrast to the strongly-adiabatic regime with $\gamma \gg A_0$, when the modulation can practically be neglected, and $T_0$ becomes equivalent to the static transmission exhibiting a Lorentzian lineshape. Equation (\ref{eqn:s_n}) provides some insight into the flat-top feature. When $A_0/\Omega \gg 1$, as is required to satisfy both $A_0 \sim \gamma$ and eq. (\ref{eqn:ad_cond}), the Bessel functions $\mathcal{J}_n(A_0/\Omega)$ have approximately comparable values for all $n < A_0/\Omega$. Thus, $s_0$ for example is given by the superposition of broad peaks ($\gamma/\Omega \gg 1$) centered at every $k \Omega$, with similar weights $\mathcal{J}_{k}^2(A_0/\Omega)$, resulting in the flat-top spectral feature.

\begin{figure}[t]
\includegraphics[width = 0.46\textwidth, trim = 0in 0in 0in 0in, clip = true]{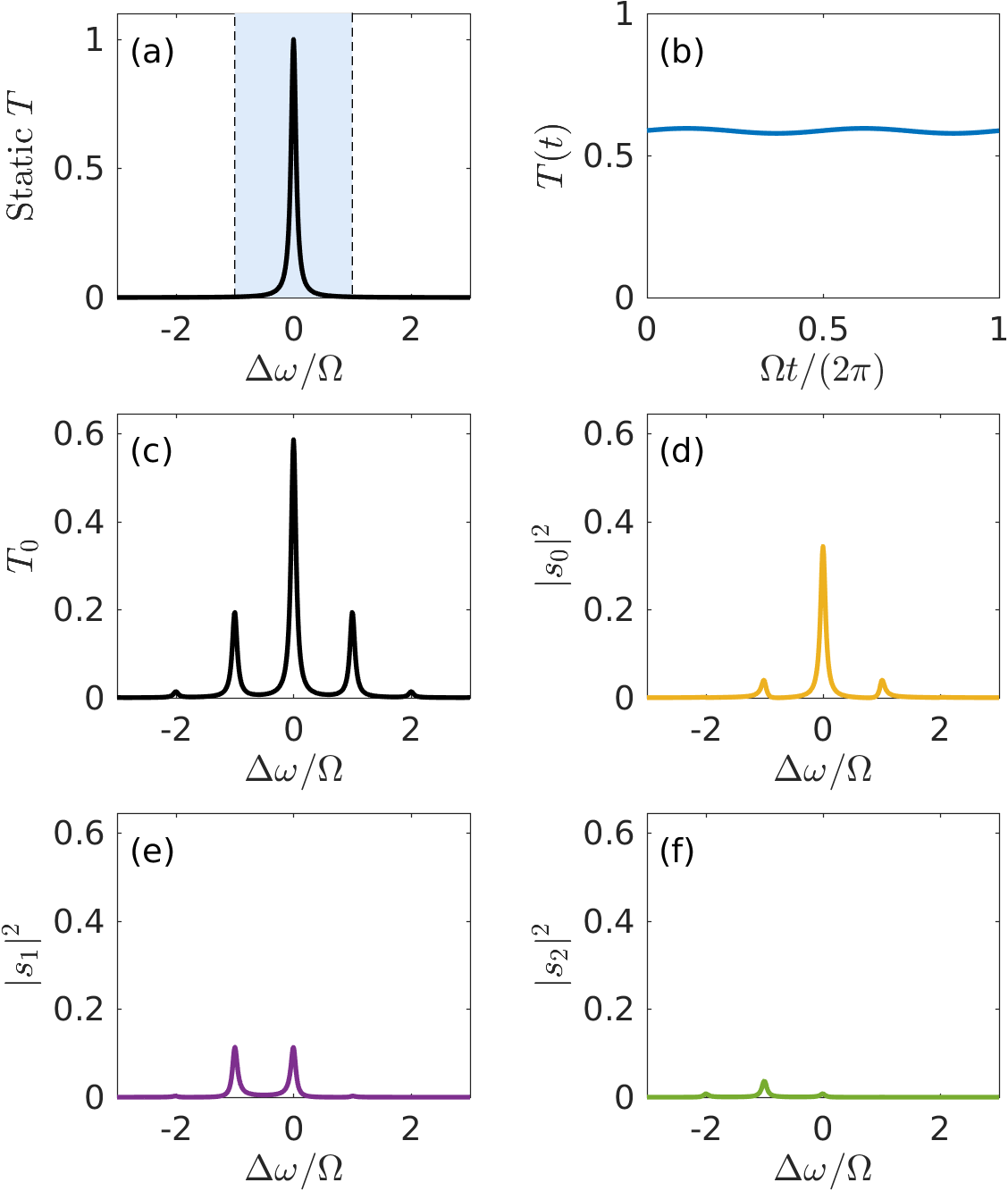}%
 \caption{Same as Fig. \ref{fig:adiabatic}, but for $\gamma/\Omega = 0.05$, $A_0/\Omega = 1$.}
\label{fig:highom}
\end{figure}

\subsection{High-frequency limit}

When the adiabatic condition is not met, the transmission has a highly non-trivial time-dependence, as described by eq. (\ref{eqn:s_n}). This expression is most easily understood in the limit of a large modulation frequency $\Omega$, or, more precisely, $\gamma/\Omega \ll 1$, when the light stays in the resonator much longer than one modulation cycle. In this case, the terms in the summation of eq. (\ref{eqn:s_n}) all become small if $\Delta \omega + k\Omega \gg \gamma$ for every $k$, while, for $\Delta \omega \approx k\Omega$ we have
\begin{equation}
s_n(\Delta \omega \approx k\Omega) = \mathcal{J}_{n+k}\left(\frac{A_0}{\Omega}\right) \mathcal{J}_k \left(\frac{A_0}{\Omega}\right) \frac{\gamma}{i\Delta \omega - ik \Omega + \gamma}.
\label{eqn:spec_approx}
\end{equation}
With this approximation, and using the identity $\sum_m \mathcal{J}_m(x) \mathcal{J}_{n+m}(y) = \mathcal{J}_n(x - y)$, we find that only the zero-th component of the transmission is non-zero, or, more precisely,
\begin{align}
 T_n(\Delta \omega \approx k\Omega) =  \mathcal{J}_k^2 \left(\frac{A_0}{\Omega}\right)\left|\frac{\gamma}{i\Delta \omega - ik \Omega + \gamma}\right|^2 \delta_{n0}.
 \label{eqn:trans_highom}
\end{align}
In other words, the transmission is time-independent in this limit, as can be intuitively expected when very fast oscillations are averaged out. Furthermore, when the input frequency is close to $\omega_0 + k\Omega$ for a given integer $k$, there is a resonance feature in the transmission spectrum with the same Lorentzian lineshape as the resonance of the static cavity, but with a magnitude scaled by $\mathcal{J}_k^2 \left(\frac{A_0}{\Omega}\right)$. This result is illustrated in Fig. \ref{fig:highom}, which shows the same plots as Fig. \ref{fig:adiabatic}, but for a DBR cavity with $\gamma/\Omega = 0.05$ and $A_0/\Omega = 1$. We see that the transmission of the modulated cavity in panel (c) is precisely given by a collection of Lorentizan peaks centered at every $k\Omega$. Each of these peaks has the same shape as the static transmission of panel (a), with height scaled by $\mathcal{J}_k^2(1)$. As expected from eq. (\ref{eqn:trans_highom}), the time-dependence shown in panel (b) for $\Delta \omega = 0$ is very close to constant. 

Interestingly, even though the transmission in this regime is time-independent, the output amplitude is \textit{not} monochromatic at $\omega_p$. Instead, it contains components at all frequencies $\omega_p + n\Omega$, with the Bessel function scaling of eq. (\ref{eqn:spec_approx}). This is illustrated in Fig. \ref{fig:highom}(d)-(f), where we plot the $s_n$ components as a function of $\Delta \omega$ for $n = 0, 1, 2$ (for the negative-$n$ counterparts, refer to eq. (\ref{eqn:s_-n})). At zero detuning, the main component in the output is the one at $\omega_p = \omega_0$. However, when $\omega_p = \omega_0 + \Omega$, the largest component is $s_{-1}$, i.e. the strongest output is at the cavity frequency $\omega_0$ and not at the input frequency $\omega_p$. This shows that there is strong frequency conversion in the transmission signal due to the modulation in this high-frequency regime.

\section{Applications}

The theoretical results presented thus far suggest a rich phenomenology of the modulated-cavity system. In this section, we illustrate several aspects that are potentially relevant for applications in photonic technologies. 

\subsection{Transmission switching}

\begin{figure}[t]
\includegraphics[width = 0.46\textwidth, trim = 0in 0in 0in 0in, clip = true]{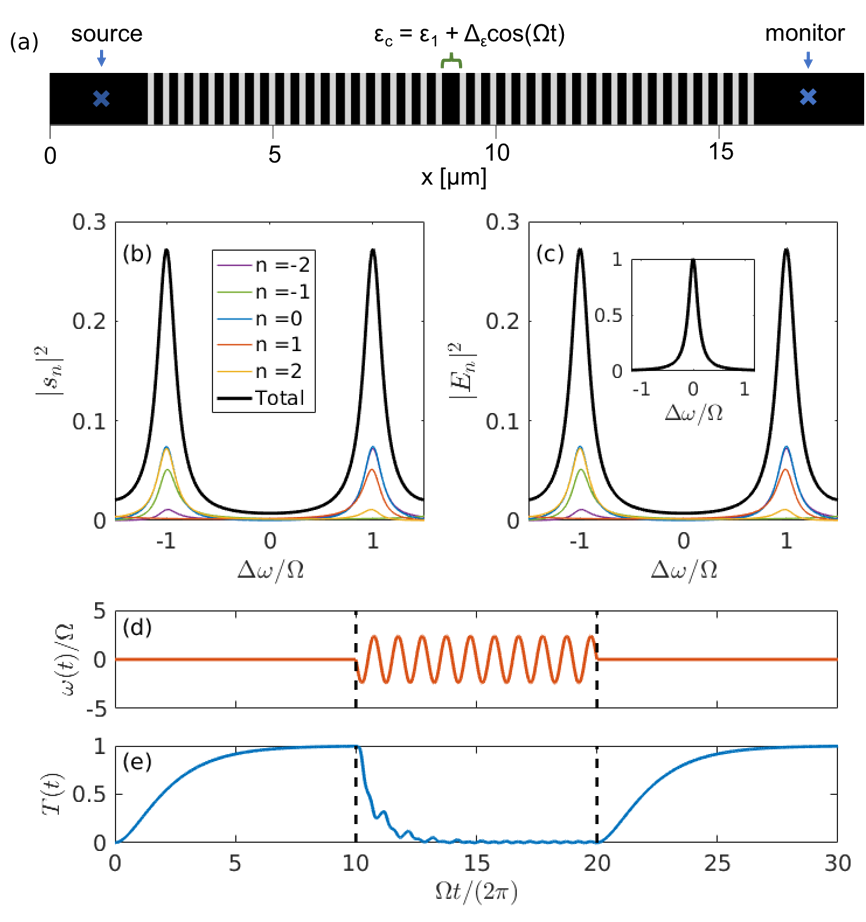}%
 \caption{(a): Schematic of the DBR for the FDFD simulation. Black and grey regions indicate permittivity $\varepsilon_1 = 4$ and $\varepsilon_2 = 8.16$, respectively. The permittivity of the central region is dynamically modulated at frequency $\Omega/(2\pi) = 5$GHz. (b): CM theory computation of the transmitted electric field amplitudes for a modulation with $A_0/\Omega = 2.405$ and $\gamma/\Omega$ as computed for the DBR of (a). (c): Same as (b), but computed with the MF-FDFD with $\Delta_\varepsilon = 2.709\times 10^{-4}$. The inset shows the transmission through the un-modulated cavity. (d)-(e): Illustration of dynamical switching with parameters as in (b), and $\Delta \omega = 0$, obtained using coupled-mode theory. The transmission (e) drops to nearly zero for the times for which the resonant frequency (d) is modulated. The black dashed lines show the time the modulation is turned on/off.}
\label{fig:switch}
\end{figure}

In the high-frequency regime, one striking consequence of eq. (\ref{eqn:trans_highom}) is that the transmission at $\Delta \omega \approx k\Omega$ goes to zero if $A_0/\Omega$ is a root of the $k$-th Bessel function. This suggests the possibility for a non-conventional switch, in which the transmission through the cavity can be tuned between zero and one through adjusting the amplitude of a \textit{time-periodic} modulation. To verify this, and more generally that the CM results presented thus far are relevant to physical implementations, we also perform a first-principle Maxwell-equations simulation using a recently-developed multi-frequency finite-difference frequency-domain (MF-FDFD) method that can incorporate a time-periodic refractive index modulation \cite{Shi2016}. Specifically, we simulate a physical DBR, schematically shown in Fig. \ref{fig:switch}(a): the cavity is composed of a central region of width $0.4$ \textmu m, with sixteen material layers of thickness $0.2$ \textmu m on each side. The relative permittivity alternates between $\varepsilon_1 = 4$ (black) and $\varepsilon_2 = 8.16$ (grey). The cavity supports a resonant mode at frequency $\omega_0/(2\pi) = 187$THz, and we include a modulation of the permittivity of the central layer, $\varepsilon_c = \varepsilon_1 + \Delta_\varepsilon \cos(\Omega t)$, with frequency $\Omega/(2\pi) = 5$GHz and $\Delta_\varepsilon = 2.71\times 10^{-4}$. A monochromatic, TE-polarized (electric field orthogonal to the $x$-axis) source excites the cavity from the left, and the transmission is recorded on the right. Due to the modulation, the electric field has a component at every side-band to the source frequency $\omega_p$, and can be written as 
\begin{equation}
E_z(\mathbf{r}, t) = \sum_n E_{zn}(\mathbf{r}) e^{i(\omega_p + n\Omega)t}.
\label{eqn:Ezn}
\end{equation}

To compare this simulation to CM theory, we first extrapolate the coupling constant $\gamma$ by fitting the transmission of the un-modulated cavity (inset of panel (c)) as a function of input frequency. In that case, $\gamma$ is simply the half-width at half-maximum of the Lorentzian peak, and is found to be $\gamma/\Omega = 0.106$. Next, we determine the dependence of the resonance frequency of the cavity on the permittivity $\varepsilon_c$ of the central region, by simulating the un-modulated structure with a slightly higher $\varepsilon_c = 4 + \varepsilon_m$. We find that, for $\varepsilon_m = 10^{-4}$, the resonant frequency changes by $\Delta_\omega/\Omega = 0.89$. This defines the relationship between permittivity change and resonant frequency change, and consequently between $\Delta_\varepsilon$ in the MF-FDFD and the $A_0$ value of CM-theory. The choice of $\Delta \varepsilon = 2.71\times 10^{-4}$ corresponds to $A_0/\Omega = 2.405$, which is a root of $\mathcal{J}_0(x)$.

Using these parameters, in panel (b) we show the CM-computed transmission including the modulation. The black curve shows the total transmission over one cycle, i.e. $T_0 = \sum |s_n|^2$. In panel (c), we plot the electric field components at the monitor position computed using the MF-FDFD. These agree perfectly with the CM result of panel (b). As expected from eq. (\ref{eqn:trans_highom}), the transmission is close to zero around $\Delta \omega = 0$ (it goes strictly to zero only in the $\gamma \rightarrow 0$ limit). The inset to panel (c) shows the transmission through the un-modulated cavity, which goes to unity at $\Delta \omega = 0$. Therefore, for a sufficiently narrow-band signal near zero detuning, the dynamic modulation switches the system from complete transmission to complete reflection.

In Fig. \ref{fig:switch}(d)-(e), we illustrate the dynamics of the switching, by turning on and then off the cavity resonant frequency modulation. The computation was carried out by numerically solving eqs. (\ref{eqn:starting1}-\ref{eqn:starting2}) using a Runge-Kutta method. At time zero, the cavity is empty ($\alpha(0) = 0$) and not modulated. The steady-state of unity transmission is reached after a time that is a few $1/\gamma$. At time $t = 10 \times (2\pi)/\Omega$, the dynamic modulation with $A_0/\Omega = 2.405$ is turned on, and the system evolves into a steady-state of near-zero transmission. At time $t = 20 \times (2\pi)/\Omega$, the modulation is switched off, and the system returns to unity transmission. It is worth emphasizing how strikingly different this regime is from the adiabatic one: throughout the modulation, the transmission is close to zero even at times at which the input is resonant with the cavity mode, and the transmission of the system at such time would be unity in the adiabatic regime. 

\subsection{Frequency conversion}

\begin{figure}
\includegraphics[width = 0.46\textwidth, trim = 0in 0in 0in 0in, clip = true]{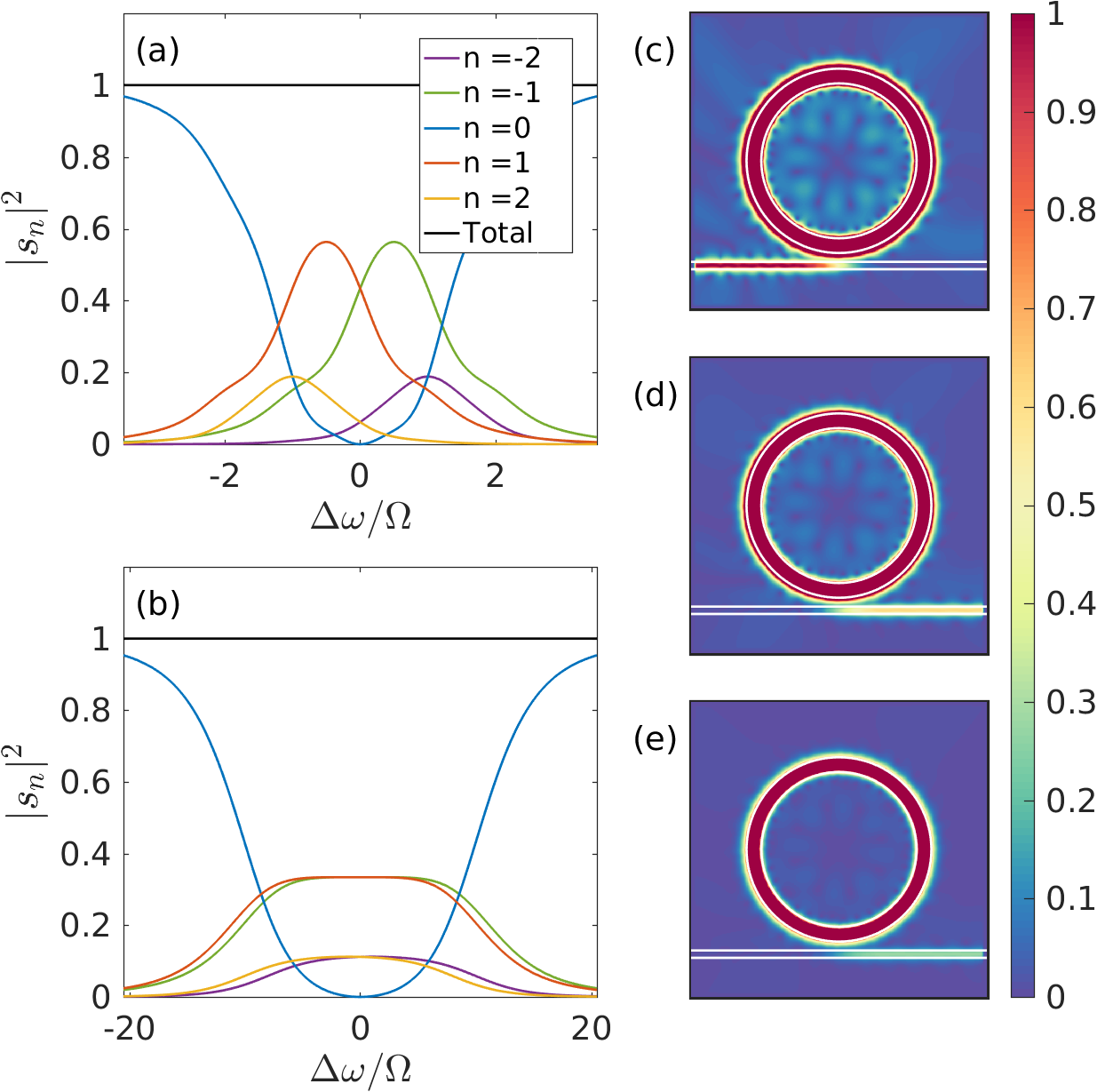}%
 \caption{(a): Power transmitted in each side-band vs. source-cavity detuning for the modulated ring cavity geometry as shown in Fig. \ref{fig1}(b), with $A_0/\Omega = 1.5$, $\gamma/\Omega = 1.5$. (b): Same as (a), for $A_0/\Omega = 10$, $\gamma/\Omega = 11.5$. (c)-(e): FDFD-computed electric field intensity in a modulated micro-ring, illustrating complete power conversion. The $|E_{z0}|^2$, $|E_{z1}|^2$, and $|E_{z2}|^2$ components are shown in (c), (d), and (e), respectively. Due to the strong field amplification inside the ring, and for better illustration, the color scale is saturated, i.e. all values larger than one are shown in red.}
\label{fig3}
\end{figure}

Next, we illustrate the possibility for complete, lossless frequency conversion in the micro-ring system. This is not possible in the DBR case, where we have $|s_0|^2 > 0$ in the whole parameter range. However, in the ring case, because of the interference between the direct and the indirect pathways -- see eq. (\ref{eqn:trans_mr}) -- we can have $s_0 = 0$ for a wide range of parameters. In such a case, the input light at $\omega_0$ is completely converted to other side bands. In fact, for $\Delta \omega = 0$, the equation $s_0(\gamma) = 0$ has a solution for every given $A_0$ larger than $A_0/\Omega \approx 1.1$. One example is shown in Fig. \ref{fig3}(a), where we plot the transmitted spectrum for $A_0/\Omega = 1.5$, $\gamma/\Omega = 1.5$, and $s_0$ indeed goes to zero around $\Delta \omega = 0$. With increasing modulation amplitude, this effect of near-complete frequency conversion can be made increasingly broad-band. In panel (b), we illustrate this for $A_0/\Omega = 10$, $\gamma/\Omega = 11.5$, and observe that $s_0 < 0.1$ in the range $|\Delta \omega|/\Omega < 6$, i.e. within a bandwidth that is an order of magnitude larger than $\Omega$. Furthermore, the frequency range for near-complete frequency conversion can be made arbitrarily large as long as the CM equations (\ref{eqn:mr1}-\ref{eqn:mr2}) are a valid description of the system. Finally, we also note that the system of panel (b) is in the adiabatic regime, which shows that non-trivial results like complete frequency conversion exist even in the adiabatic limit.

We verify these results using the multi-frequency FDFD method of Ref. \cite{Shi2016} to simulate a ring cavity side-coupled to a waveguide (Fig. \ref{fig3}(c)-(e)). The ring and waveguide are assumed to be silicon with permittivity $\varepsilon_2 = 12$. The surrounding material is air having a permittivity of $\varepsilon_1 = 1$. The waveguide has a width of $0.2$ \textmu m, while the ring waveguide has a width of $0.4$ \textmu m and an outer radius of $2.5$ \textmu m. The system is excited from the left by a source of frequency $\omega_p$, located in the center of the waveguide. The ring has an $E_z$-polarized mode resonant at $\omega_0/(2\pi) = 200.4$THz. A CM-fit of the transmission allowed us to extrapolate the ring-waveguide coupling to be $\gamma/(2\pi) = 9.5$GHz, and also suggested that extra radiative losses (radiation not coupled into the waveguide) were present as characterized by an intrinsic loss rate $\gamma_L/(2\pi) = 0.55$GHz. The effect of such an extra loss is easily incorporated in eq. (\ref{eqn:trans_mr}), by replacing $\gamma/2$ with $(\gamma + \gamma_L)/2$ in the denominator. With these coupled-mode theory predicts that complete frequency conversion occurs for $A_0/\Omega = 1.7$, when, at $\Delta \omega = 0$, we have $|s_0|^2 = 0$, $|s_1|^2 = 0.377$, $|s_2|^2 = 0.063$, $|s_3|^2 = 0.005$.

A simulation of the static cavity with $\varepsilon_r = \varepsilon_2 + \varepsilon_m$, with $\varepsilon_m = 10^{-3}$, resulted in a shift of the resonant frequency by $8.32$GHz. This allows us, as in the DBR case, to map the modulation amplitude $A_0$ of coupled-mode theory to the amplitude of the permittivity modulation. We modulate the entire ring such that $\varepsilon_r = \varepsilon_2 + \Delta_\varepsilon \cos(\Omega t)$, with $\Omega/(2\pi) = 5$GHz. We find complete conversion for $\Delta_\varepsilon = 1.2 \times 10^{-3}$, which implies a modulation amplitude of $A_0/\Omega = 2.0$, which is slightly larger than the required modulation amplitude predicted by the coupled-mode theory. In Fig. \ref{fig3}(c)-(e), we plot the components $|E_{zn}|^2$ of eq. (\ref{eqn:Ezn}), for $n = 0, 1, 2$, normalized to  $|E_{z0}|^2$ at the source position. As can be seen, no power is transmitted at the source frequency -- all the power is instead completely converted to the other side-bands. The normalized transmitted power components in the center of the waveguide are $|E_0|^2 = 0$, $|E_1|^2 = 0.373$, $|E_1|^2 = 0.079$, $|E_2|^2 = 0.008$, which compare very well to the CM-computed $s_n$ components. We note that, for $\Delta \omega = 0$ as is the case here, the $|E_{\pm n}|^2$ components are equal for all $n$, and that the total transmitted power is $\sum_n |E_n|^2 = 0.92$. This is less than one because of the non-zero loss $\gamma_L$. Importantly, the system does \textit{not} need to be operated around critical coupling, and the losses can in principle be arbitrarily small by increasing the $\gamma/\gamma_L$ ratio.

\subsection{Signal optimization}

\begin{figure}
\includegraphics[width = 0.46\textwidth, trim = 0in 0in 0in 0in, clip = true]{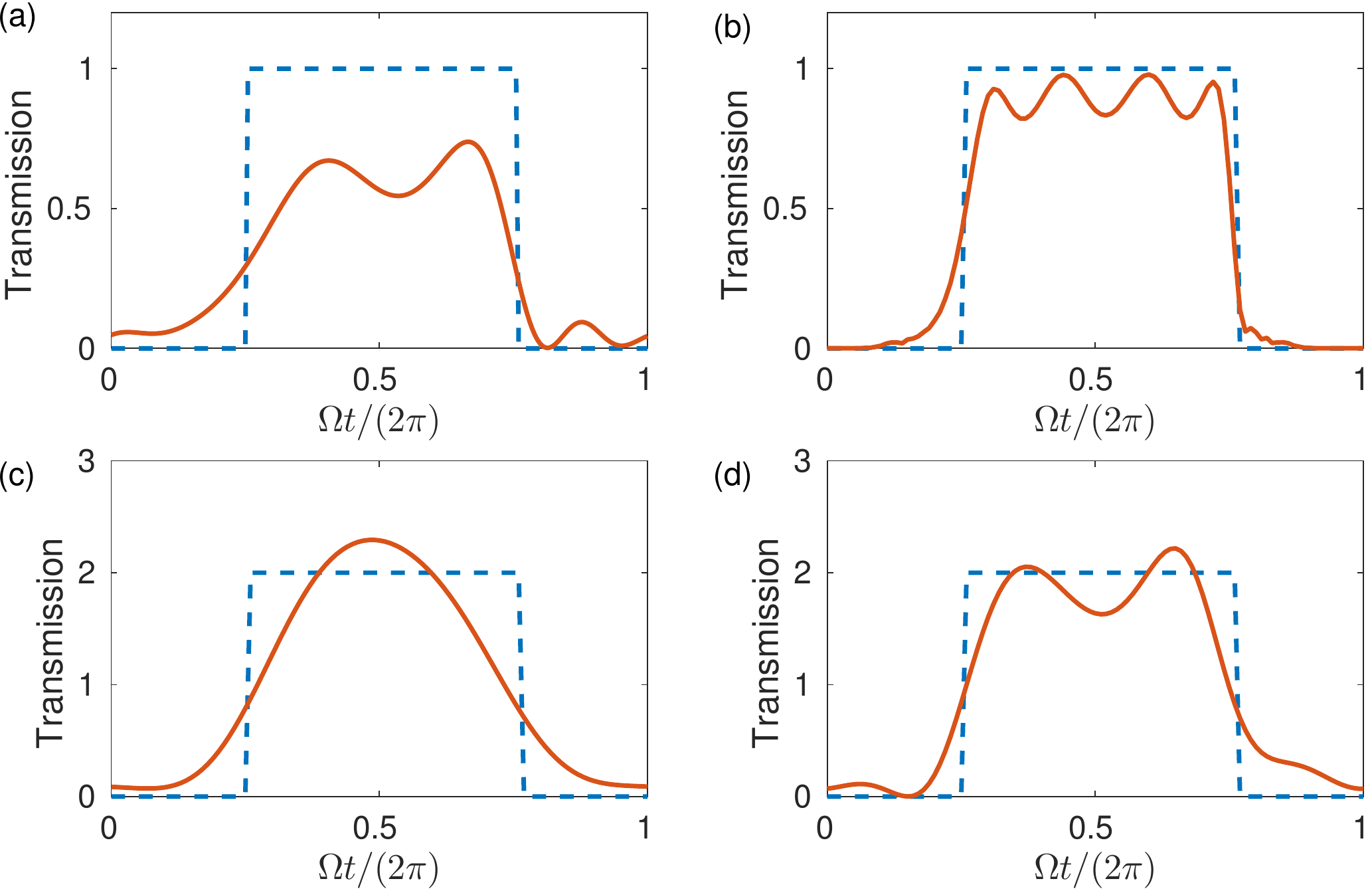}%
 \caption{Optimizing the time-dependent transmission of a cavity. Blue dashed lines show the target signal, red lines show the closest fit for (a): DBR cavity modulated cosinoidally at a frequency $\Omega$; (b): DBR with a second modulation at $2\Omega$; (c): ring cavity modulated cosinoidally at a frequency $\Omega$; (d): ring with a second modulation at $2\Omega$.}
\label{fig:opt}
\end{figure}

Finally, we also demonstrate how the exact steady-state solution obtained here can be used to engineer a particular non-trivial transmission signal. As a specific example, we target a transmission $T_t(t)$ shown as dashed lines in Fig. \ref{fig:opt}, featuring a periodic step-function switching on/off the transmission. Generating such a profile can be beneficial for e.g. optical clock distribution \cite{Clymer1986, Mule2002} or optical sampling \cite{Schmidt-Langhorst2005}. 

To achieve the target transmission, we perform an optimization seeking to maximize the overlap between the CM-computed and the target transmission. We define $\tilde{T}_{t, n}$ as the normalized Fourier components of the target transmission, and $\tilde{T}_n$ as the  normalized components computed through eq. (\ref{eqn:trans}). Furthermore, we allow for an arbitrary phase detuning $\phi_t$, such that the objective function reads 
\begin{equation}
f(\mathbf{x}) = \sum_n \tilde{T}^*_{t,n} e^{i\phi_t} \tilde{T}_n, 
\end{equation}
The vector of parameters $\mathbf{x}$ includes $\phi_t$ as well as the free parameters of eq. (\ref{eqn:s_n}). For the modulation $\omega(t) = A_0 \cos(\Omega t)$, then, we have $\mathbf{x} = (\gamma$, $A_0$, $\Delta \omega, \phi_t)$. Using a steepest descent method with multiple starting points, we reach an optimal $\mathbf{x}$ the maximizes $f(\mathbf{x})$.  The optimal overlap is found for  $\Delta \omega/\Omega = 2.7$, $A_0/\Omega = 4.5$, $\gamma/\Omega = 1.25$, and the transmitted signal is plotted in Fig. \ref{fig:opt}(a).

To improve the optimization, we expand the number of parameters by considering a second modulation at $2\Omega$ acting on the cavity, such that $\omega(t) = A_1 \cos(\Omega t) + A_2 \cos(2\Omega t + \phi_m)$ . The side-band components of the transmitted amplitude in this case are given by eq. (\ref{eqn:sn_2m}). With these new parameters, the best overlap with the target transmission is reached for $\Delta \omega/\Omega = 33.6$, $A_1/\Omega = 52.6$, $\gamma/\Omega = 8.3$, $A_2/\Omega = 23.0$, $\phi_m = 3.1$. The corresponding transmission signal is shown in Fig. \ref{fig:opt}(b), and comes very close to the target step-function. We note that such a signal with features much shaper than the time-scale given by $\Omega$ is highly non-trivial, and it is striking that it can be obtained from a system as simple as ours. For completeness, we also show the results of the same optimizations performed for the micro-ring cavity. In panel (c), using a single cosine modulation, the best parameters were found to be $\Delta \omega/\Omega = 0.09$, $A_0/\Omega = 1.1$, $\gamma/\Omega = 0.1$. In panel (d), using a second modulation, the optimal parameters are $\Delta \omega/\Omega = 1.1$, $A_1/\Omega = 0.92$, $\gamma/\Omega = 0.47$, $A_2/\Omega = 1.35$, $\phi_m = 1.7$. The inclusion of the $2 \Omega$ components again leads to better overlap with the target. 

\section{Conclusion}

In conclusion, we have presented a detailed study of the steady-state dynamics of an optical resonator coupling with one or two input/output ports, and subject to a periodic modulation of the resonance frequency and a continuous-wave input. The exact solution that we have derived provides intuition in and beyond the adiabatic limit, and suggests interesting features of the phenomenology of this system. These include dynamic decoupling from the source, as well as the potential for complete, lossless frequency conversion within a large bandwidth around the resonant frequency. These results can lead to novel functionalities of electro-optic modulators in the field of communications, and may also be relevant to frequency comb generation \cite{Ye1997, Griffith2015} and optomechanical systems \cite{Hafezi2012, Johnson2006}. The analytic result also allows for a quick yet exhaustive exploration of the parameter space, and can be used to optimize the transmission towards a given target. In short, this conceptually simple system, which is a basic building block of on-chip photonic technologies, was found to show very rich physics that goes well beyond the applications that it has found thus far. 

\vspace{0.5cm}

\section{Acknowledgement}

This work was supported by the Swiss National Science Foundation through Project N\textsuperscript{\underline{o}} P2ELP2\_165174, and the US Air Force Office of Scientific Research Grant No. FA9550-17-1-0002. 


%

\end{document}